# Raman Spectroscopy Study of α-, β-, γ-Na$_x$CoO$_2$ and γ-(Ca,Sr)$_x$CoO$_2$

H.X. Yang, Y. Xia, Y.G. Shi, H.F. Tian, R.J. Xiao, X. Liu, Y.L. Liu and J.Q. Li*

Beijing National Laboratory for Condensed Matter Physics, Institute of Physics, Chinese Academy of Sciences, Beijing 100080, China

Raman spectroscopy measurements have been performed on α-, β-, and γ-Na$_x$CoO$_2$ phases differing in their stacking of CoO$_6$ octahedra along the *c*-axis direction. The results demonstrate that, in general, there are five Raman active phonons for γ-Na$_{0.75}$CoO$_2$, two Raman active phonons for α-NaCoO$_2$, and three Raman peaks for β-Na$_{0.6}$CoO$_2$. We have also performed Raman scattering measurements on several γ-(Ca,Sr)$_x$CoO$_2$ (0.15 ≤ x ≤ 0.40) samples which show well-defined intercalated Ca/Sr-ordering. The experimental data show that the intercalated cation ordering could result in visible alterations on Raman spectral structures. The observations of the spectral changes along with the variation of the CoO$_6$ stacking, as well as the intercalated Sr/Ca ordering suggest that the interlayer interaction plays an important role for understanding the phonon properties in this layered system.

1. Introduction

   Layered $Na_xCoO_2$ materials have been extensively investigated in recent years due to a notably large thermoelectric power coexisting with superconductivity and complex charge-ordering (CO) transitions [1-3]. In order to understand the lattice dynamics in this layered system, Raman scattering [4-9], infrared absorption [10], and neutron scattering experiments [11] have been carried out in numerous laboratories and under different experimental conditions. Theoretical analysis based on lattice shell model and first principle calculations have also been performed for typical $Na_xCoO_2$ materials with x = 1.0, 0.74, 0.5 and 0.3 [5, 12], however, much controversy still remains in related published results [4-13]. J.F. Qu et al. observed three phonon modes in the experimental spectrum of $Na_{0.7}CoO_2$ [13], in sharp contrast with our previous reported data which clearly demonstrated five Raman-activated modes for γ- $Na_{0.75}CoO_2$ [4]. Moreover, remarkable spectral differences also exist in the charge ordered $Na_{0.5}CoO_2$ material as reported in refs. 5, 6, 9 and 13. In our recent study, it is noted that the layered $Na_xCoO_2$ materials actually have a rich variety of structural features that could have visible effects on physical properties, such as, structural modulation from Na-ordering, and sample (surface) instability in the air or water [4, 6, 14]. There are three distinctive structural series of $Na_xCoO_2$ materials that differ in their stacking of $CoO_6$ octahedra along the *c*-axis direction, the so called α-, β-, and γ- phases respectively. The distance between two $CoO_2$ layers are almost the same for all three phases, each phase having its specific structural properties depend on Na concentration [15-21]. It is also noted that structural alternations in $Na_xCoO_2$ can also be caused by the post-chemical treatments as commonly adopted for preparing the $Na_xCoO_2$ sample with lower Na concentrations. For instance, our recent experimental results revealed that α-$Na_xCoO_2$ materials undergo a clear phase transition from the hexagonal to the β-phase-like monoclinic structure along with the Na deintercalation [14]. Actually, the structural stability along with the decrease of Na concentration has not been very well understood for this layered system, especially, for α- and β- $Na_xCoO_2$. Therefore, special precautions have to be taken for the samples used in experimental measurements. Cation ordering, commonly existing in layered $Na_xCoO_2$ materials, is another notable issue concerned in structural measurements and analysis of physical properties. For instance, certain experimental and theoretical investigations have paid special attention to intercalated Na ordering among $CoO_2$ sheets [6, 22], where the $3^{1/2} \times 3^{1/2}$ superstructure was expected to have marked effects on the magnetic ordering in the layered system [23]. Recently, we have synthesized a series of γ-$Ca_xCoO_2$ (0.15 ≤ x ≤ 0.4) and γ-$Sr_xCoO_2$ (0.15 ≤ x ≤ 0.4) by means of an ion exchange reaction. Structural analysis indicates that the average structures of these materials are isomorphic with the γ-$Na_xCoO_2$ phase. The well-defined cation ordering has been characterized by TEM observations [24, 25]. Moreover, $M_xCoO_2$ (M = Sr, Ca) compounds are found to be much more stable in the air than their analogous $Na_xCoO_2$ materials, thus $(Sr,Ca)_xCoO_2$ are more suitable for Raman observations [24, 25]. In the present paper, we will report on the Raman scattering measurements on the parent phases of α-, β-, γ-$Na_xCoO_2$ and on typical samples of γ-$(Sr,Ca)_xCoO_2$ (0.15 ≤ x ≤ 0.40). Certain spectral features have been also discussed in comparison with the theoretical analysis based on the group factor theory [26].

2. Experimental

Polycrystalline samples of $\alpha$-NaCoO$_2$, $\beta$-Na$_{0.6}$CoO$_2$, $\gamma$-Na$_{0.75}$CoO$_2$ and (Sr,Ca)$_x$CoO$_2$ (0.15 ≤ x ≤ 0.40) materials were used in our Raman scattering study. The $\gamma$-Na$_{0.75}$CoO$_2$ phase was synthesized following the procedure as described in Ref [17], while $\alpha$-NaCoO$_2$ and $\beta$-Na$_{0.6}$CoO$_2$ were prepared by conventional solid-state reactions at relatively lower temperatures of 500°C to 550°C [14]. In the synthesis of $\alpha$-NaCoO$_2$ materials, powdered cobalt (Co) metal (99.5%) and 10% excess anhydrous NaOH pellets (Aldrich) (Na:Co = 1.1:1) were ground together under inert atmosphere and placed in an alumina boat under flowing O$_2$ for approximately 6 days at 500°C with one intermediate grinding. For the synthesis of $\beta$-Na$_{0.6}$CoO$_2$ material, the Co powder and NaOH pellets were mixed in a molar ratio of Na:Co = 0.7:1, ground together under flowing N$_2$, and then reacted under flowing O$_2$ at 550°C for approximately 5 days with one intermediate grinding. Polycrystalline materials with nominal compositions of Sr$_x$CoO$_2$ and Ca$_x$CoO$_2$ (0.15 ≤ x ≤ 0.40) were prepared by a low-temperature ion exchange reaction using a $\gamma$-Na$_x$CoO$_2$ (0.33 ≤ x ≤ 0.8) precursor prepared by conventional solid-state reaction or by sodium deintercalation of Na$_{0.75}$CoO$_2$ [27]. The ion exchange process was carried out using the modified Cushing-Wiley method [28]. X-ray diffraction (XRD) measurements were carried out with a diffractometer in the Bragg-Brentano geometry using Cu K-$\alpha$ radiation. The compositions of all materials have been measured by an inductively coupled plasma (ICP) analysis technique. XRD results show that Sr/Ca intercalated samples made by Cushing-Wiley method have very similar structure with their $\gamma$-Na$_x$CoO$_2$ precursors, in good agreement with previous publications [24, 25, 28].

Specimens for transmission-electron microscopy (TEM) observations were polished mechanically with a Gatan polisher to a thickness of around 50μm and then ion-milled by a Gatan-691 PIPS ion miller. The TEM measurements were performed on the Tecnai F20 (200 kV). Raman spectra were collected using a back-scattering geometry at room temperature with a Jobin-Yvon T64000 triple spectrometer equipped with a cooled charge-couple device. In the spectrometer, an objective of 100X-magnification was used to focus the laser beam on the sample surface and collect the scattered light. The excitation wavelength of 514.5 nm of an $Ar^+$ ion laser was used in our experiments. The laser power at the focus spot of 2-3 μm in diameter was kept below 1 mW to prevent laser-induced damage to the samples. For storage, the samples were rapidly put in a vacuum to avoid surface contamination and water adsorption. For facilitating the comparison, The Raman spectra presented in figure 2 and 4 have been stacked by some arb. units with respect to each other without being multiplied by any scale factor.

3. Results and discussion

Figure 1(a) shows the XRD patterns for the α-$NaCoO_2$, β-$Na_{0.6}CoO_2$ and γ-$Na_{0.75}CoO_2$ phases. These phases differ crystallographically in the stacking patterns of $CoO_6$ octahedra along the *c*-axis direction [15-21]. In the α-$Na_xCoO_2$ system, the parent phase with a nominal composition of α-$NaCoO_2$ has a hexagonal cell with lattice parameters of a = 2.89 Å, c = 15.59 Å ($R\bar{3}m$ space group) [20]; the parent sample β-$Na_{0.6}CoO_2$ has a monoclinic structure with a C2/m space group and lattice parameters a =

4.90 Å, b =2.83 Å, c = 5.71 Å, β = 106.180° [21]; The parent material of γ-$Na_{0.75}CoO_2$ has a hexagonal cell with lattice parameters a = 2.84 Å and c = 10.80Å ($P6_3$/mmc space group) [11]. In recent studies, the γ-$Na_xCoO_2$ (0.15 < x < 0.75) materials were found to show a rich variety of significant physical properties [1-3], such as, a notably large thermoelectric power, superconductivity under water intercalation, and CO transitions for x~0.5.

Figure 1(b) shows the layered structural features of α-$NaCoO_2$, β-$Na_{0.6}CoO_2$ and γ-$Na_xCoO_2$. The α-$NaCoO_2$ contains three $CoO_2$ layers in one unit cell in which Na occupies the unique 3a (0,0,0) site, Co occupies the 3b (0,0,1/2) site, and O occupies the 6c (0,0,z) site [20]. The β-$Na_xCoO_2$ has a C2/m symmetry, its unit cell contains only one $CoO_2$ layer, Na occupies the 2a (0,0,0) site, Co occupies the 4i (x,0,z) site, and O occupies the 8j (x,y,z) site [21]. The γ-$Na_{0.75}CoO_2$ material contains two $CoO_2$ layers in one unit cell in which Co occupies the 2a (0,0,0) and O occupies the 4f (1/3,2/3,z) site, while the Na atoms in this structure could occupy two distinct sites within a given plane denoted with Wyckoff indices of 2b($Na_1$) and 2d($Na_2$) [11]. The occupation ratio on either $Na_1$ or $Na_2$ site depends on the Na content.

Based on this structural information, we have made a brief theoretical factor-group analysis for the Raman active modes in the α-, β- and γ-$Na_xCoO_2$ phases, based on the method as reported in Ref [26]. It is demonstrated that there are two Raman active phonon modes $A_{1g}$+$E_g$ for the α-$NaCoO_2$ phase, both connected with O motions; there are nine Raman active phonon modes: 5$A_g$+4$B_g$ for β-$Na_{0.6}CoO_2$, $A_g$ and $B_g$ modes may connect with both Na and O motions. The γ-$Na_xCoO_2$ phase has five Raman active phonon modes: $A_{1g}$ + $E_{1g}$ + 3$E_{2g}$, the $A_{1g}$ and $E_{1g}$ modes involve motions of the oxygen atoms only, $E_{2g}$ modes may connect with both Na and O motions while Co motions are not Raman active as discussed previously [4].

Fig. 2 shows the experimental Raman spectra for three typical phases. According to the conductivity measurements for these phases, the α-NaCoO$_2$ phase is a semiconductor at room temperature while β-Na$_{0.6}$CoO$_2$ and γ-Na$_x$CoO$_2$ phases are metallic [29]. An increase in conductivity reduces the optical skin depth of the incident laser beam, and results in a decrease of the signal/noise ratio. Hence, under the same experimental conditions the α-NaCoO$_2$ has a relatively higher signal/noise ratio. The experimental spectrum of α-NaCoO$_2$ contains two clear peaks found at 486.5 and 586.4cm$^{-1}$, respectively. We interpret theses peaks as the E$_g$ (486.5cm$^{-1}$) and A$_{1g}$ (586.4cm$^{-1}$) active modes in accordance with theoretical analysis. The *A$_{1g}$* and *E$_g$* modes involve atomic motions from oxygen atoms only: the *E$_g$* represents the in-plane vibration, and the *A$_{1g}$* represents the out-of-plane movement. The energy of the *A$_{1g}$* mode strongly depends on the occupancy of the Na layer which divides the CoO$_6$ octahedral in the c-axis direction. It is worth mentioning that our experimental spectra for the α-Na$_x$CoO$_2$ phase shows noticeable similarities with results reported in Ref [5] and [9] obtained from Na$_x$CoO$_2$ single crystals. Hence, it is possible that the above referenced crystals (or crystal surfaces) crystallize in an α-phase structure rather than a γ-phase structure. Experimental spectra of β-Na$_{0.6}$CoO$_2$ in general contain three strong peaks at around 461, 577, and 691 cm$^{-1}$ respectively (see Fig. 2). On the other hand, our theoretical factor-group analysis suggests that nine active modes (5A$_g$+4B$_g$) are possibly visible; this discrepancy could be caused by the overlap of Raman peaks. There are five Raman active modes that can be identified: *A$_{1g}$* at 673.3cm$^{-1}$, *E$_{1g}$* at

188.5cm$^{-1}$, and $E_{2g}$ at 470.4cm$^{-1}$, 510.7cm$^{-1}$, 605.8 cm$^{-1}$ and 673.3 cm$^{-1}$ for γ-Na$_{0.75}$CoO$_2$, which agree well with our former detailed analysis on single crystals [4].

It is noted that the intercalations of Na, Sr, and Ca atoms or H$_2$O molecules could make the local structure of this layered system much more complex. The intercalated atoms can be random with high mobility or crystallized in a variety of ordered states [6, 22]. Certain ordered states were demonstrated to have notable effects on physical properties. For instance, Na atoms in Na$_{0.5}$CoO$_2$ crystallize in an orderly, well-defined zigzag pattern yielding an orthorhombic structure in which low temperature charge ordering is observed [6]. The Raman spectroscopy is sensitive to the ordered structures, and can be used effectively to characterize the degree and type of cation order in perovskite oxides [29]. Recently, experimental and theoretical investigations have paid special attention to intercalated cation/vacancy ordering in the layered cobalt oxides [6, 22]. First principle calculations demonstrated that a variety of possible ordered states can be stable at different levels of intercalated cation content, such as $3^{1/2}a \times 3^{1/2}a$ superstructure at x = 1/3 (or 2/3) and other ordered states at x = 1/2, 1/4, 1/5, etc. [22]. Previously, several superstructures in Li$_x$CoO$_2$ and Na$_x$CoO$_2$ materials have been demonstrated to have cation ordering [6, 30]. However, it is also noted that the electron beam radiation during TEM observation could severely alter the cation arrangements among the CO$_2$ sheets due to the high mobility of either Na or Li ions in this layered structure [6, 30].

Figs. 3(a) and (b) show respectively the electron-diffraction patterns and high-resolution TEM images for the Sr$_{0.35}$CoO$_2$. The most notable structural phenomenon revealed in the electron diffraction pattern of Fig. 3(a) is the appearance of systematic weak reflection spots in addition to the main diffraction spots indexed perfectly on the known hexagonal structure. The superstructure spots in the present case can be well characterized by an in-plane wave vector **q** = (1/3, 1/3, 0) which yields a $3^{1/2}a \times 3^{1/2}a$ super-cell within the basic *a-b* plane. The superlattice spots on the **a\*-b\*** plane in general are very sharp, indicating a relatively long coherent length (>30nm) of the ordered state. Fig. 3(b) shows a [001] zone-axis HRTEM image for $Sr_{0.35}CoO_2$, illustrating the atomic structure for this superstructure. The image was obtained from a thin region of the crystal near the Scherzer defocus value ($\approx$-60nm). The metal atom (Co) positions are therefore recognizable as dark dots. In this image, the hexagonal superstructure is clearly seen in the relative thick area as illustrated by a hexagonal supercell.

Raman scattering measurements on $\gamma$-$Sr_xCoO_2$ ($0.15 \leq x \leq 0.40$) materials have revealed notable changes of the Raman spectrum along with the appearance of cation ordered states. Fig. 4(a) shows a series of Raman spectra taken from the $\gamma$-$Sr_xCoO_2$ materials with x = 0.15, 0.25, 0.35 and 0.40. These spectra have markedly different signal/noise ratio caused by the variation of the conductivity that depend directly on the Sr concentration. Measurements of resistivity showed that $Sr_xCoO_2$ in general are insulators for samples with low strontium content (x < 0.2) and transforms to metals for x > 0.35 [31]. It is noted that the Raman active modes for x = 0.15 and 0.25 show five clear active modes (with limited shifts) notably similar with the data from $\gamma$-$Na_{0.75}CoO_2$ as shown in Fig.2. Actually, the intercalated Sr atoms in $\gamma$-$Sr_xCoO_2$ samples with x < 0.25 are rather random as observed for Na atoms in $\gamma$-$Na_{0.75}CoO_2$. Certain notable spectral features were observed in the data obtained from $Sr_{0.35}CoO_2$ which contains a well defined $3^{1/2}a \times 3^{1/2}a$ superstructure, in addition to the remarkable shifts of the five typical peaks for the γ-$Sr_{0.15}CoO_2$ phase. For instance, the notable new mode at ~148 cm$^{-1}$ are possibly in correlation with Brillouine zone folding in superstructure phase. On the other hand, it is noted that the alterations of phonon width and lineshape are also visible for certain Raman active modes, these changes could be partially caused by the modifications of electronic structure; the defects and locally distortion could also result in the broadening of the Raman modes in this system.

In order to further understand the effects of intercalated cation ordering in the Raman spectra of this layered system, we have also performed measurements on the γ-$Ca_xCoO_2$ materials. This system, being similar to $Sr_xCoO_2$ and $Na_xCoO_2$, contains a rich variety of structural phenomena, such as Ca ordering and phase separation [25]. Fig. 4(b) displays the Raman spectra from the γ-$Ca_xCoO_2$ (0.15 ≤ x ≤ 0.40) materials, illustrating clear alternations of the relative intensities and frequencies of the Raman peaks with the increase of Ca content. These facts, in combination with results from $Sr_xCoO_2$ and $Na_xCoO_2$, suggest that the lattice dynamics are somewhat sensitive to the variation of the cation ordering in the present system. The γ-$Ca_{0.15}CoO_2$ material has five active modes identified respectively as $A_{1g}$ at 691.1cm$^{-1}$, $E_{1g}$ at 196.6cm$^{-1}$, and $E_{2g}$ at 483.3cm$^{-1}$, 525.1cm$^{-1}$, and 621.8 cm$^{-1}$ that are comparable with the active modes of γ-$Sr_{0.15}CoO_2$ as discussed above. Moreover, it is remarkable that all the Raman active modes show significant shifts towards higher frequencies in comparison with those in the γ-$Na_{0.75}CoO_2$ sample, especially $A_{1g}$ and $E_{1g}$ peaks which are essentially connected with the O vibrations. This could be due to the larger electrostatic interaction between divalent ions and the negative $CoO_2$ layer in γ-$(Ca,Sr)_xCoO_2$ compounds.

Systematic structural analysis on $\gamma$-$Ca_xCoO_2$ suggests that there are two well-defined cation ordered states corresponding respectively to the orthorhombic superstructure at around x = 1/2 and the $3^{1/2}a \times 3^{1/2}a$ superstructure at around x = 1/3 in this kind of system. Multiple ordered states, phase separation, and incommensurate structural modulations commonly appear in materials with 0.3 < x < 0.5 [24]. Although $\gamma$-$Ca_{0.3}CoO_2$ in general contains clear $3^{1/2} \times 3^{1/2}$ superstructure as revealed in our TEM observations [24], its Raman spectrum, in contrast to $Sr_{0.35}CoO_2$, does not shows the additional new mood at around 148$^{-1}$. This difference maybe caused by coexistence of complex multiple superstructures in $\gamma$-$Ca_{0.3}CoO_2$. Actually, the $3^{1/2} \times 3^{1/2}$ superstructure in $\gamma$-$Ca_{0.3}CoO_2$ is less well-defined in comparison with $Sr_{0.35}CoO_2$; our high resolution TEM observations demonstrated that incommensurate modulations and phase separation often appears in the $\gamma$-$Ca_{0.3}CoO_2$ samples [24]. Hence, the low frequency mode proposed in connection with Brillouine zone folding is more difficult to be detected in Raman scattering. On the other hand, some alternations that might relate to the $3^{1/2} \times 3^{1/2}$ superstructure still can be recognized in the Raman spectrum of $\gamma$-$Ca_{0.3}CoO_2$ (as shown in Fig.4b). Further analysis of the $\gamma$-$Ca_{0.3}CoO_2$ spectrum shows that the $A_{1g}$ mode at 691cm$^{-1}$ shifts slightly to a higher frequency and the $E_{2g}$ mode at 483 cm$^{-1}$ shifts to a lower frequency in comparison with those for the x = 0.15 sample. Moreover, the two modes that appear at 560 and 595 cm$^{-1}$ are possibly the remaining $E_{2g}$ modes under significant shift, or new modes arising from local structural ordered states in the superstructure phase. The spectra of $Ca_{0.35}CoO_2$ and $Ca_{0.4}CoO_2$ shows certain combined features of $Ca_{0.15}CoO_2$ and $Ca_{0.30}CoO_2$, indicating the presence of local structural inhomogeneities in the samples studied, which agrees well with the structural analysis [24].

As mentioned in the introduction part, that the Raman scattering measurements on $Na_xCoO_2$ materials (polycrystalline samples or single crystals) performed in several laboratories show apparently different spectral features [4 - 9]. Here we make a brief discussion on the samples used for experimental investigations. In our previous publication, we identified that five active phonons generally appear in the Raman spectra of $\gamma$-$Na_xCoO_2$ by using $\gamma$-$Na_{0.7}CoO_2$ single crystals grown by the flux method [4]. On the other hand, in the paper of P. Lemmens *et al*, Raman spectroscopy has been measured on $Na_xCoO_2$ single crystals, with x = 0.5, 0.83, and 1.0 [9]; all spectra show two strong phonon modes at around 580 cm$^{-1}$ and 480 cm$^{-1}$. The $NaCoO_2$ (x = 1.0, 0.83) samples used by P. Lemmens are made via floating zone method, and other samples were made by Na-deintercalation using post-chemical treatments. The spectra of x = 1.0 sample is very similar with our data obtained from the $\alpha$-$NaCoO_2$ phase in this paper. It is well known [17] that $Na_xCoO_2$ samples with x > 0.80 generally dominate by the $\alpha$-phase rather than $\gamma$-phase. Na-deintercalation can also result in certain phase transitions that are still not well understood [14]. These facts suggest that the notable differences among the spectra from different groups possibly caused by the structural discrepancy of the samples made by different methods.

The structural transitions appear during post-chemical treatments of $Na_xCoO_2$ samples are another kind of significant issues for understanding their physical properties, e.g. Na-deintercalation and $H_2O$ intercalation could result in clear lattice instability and complex defects, especially in single crystals. We have carefully checked the microstructure of the single crystalline samples of $Na_{0.5}CoO_2$ and $Na_{0.3}CoO_2$ prepared by Na-deintercalation and it was observed that stacking faults and intergrowth (coexistence) of different phases occasionally appear in the samples. Fig. 5 shows the [010] zone-axis HRTEM image and the selected-area electron diffraction pattern taken from a $Na_{0.5}CoO_2$ single crystal prepared using Na-deintercalation. As we mentioned in above context, one unit cell of $β-Na_xCoO_2$ contains only one $CoO_2$ layer with the lattice parameter of c=5.71 Å and the $γ-Na_xCoO_2$ phase contains two $CoO_2$ layers in one unit cell with c =10.80 Å. These features are clearly recognizable from the periodicities of the lattice fringes shown in Fig. 5 (as indicated by arrow), in which the top region is governed mainly by the γ-phase and the bottom area is the β-phase. The reflections in the diffraction pattern exhibit weak diffuse spots streaking along the c*- axis direction, indicating the presence of stacking faults and other planar defects in this system. It is interesting to point out that certain reported Raman scattering results for $Na_xCoO_2$, such as the spectra in Ref [5], can be interpreted by a combination of α, β, and γ phases in the range of 400 to 800 $cm^{-1}$ [5].

4. Conclusions

The Raman scattering spectra obtained from the α, β, γ -$Na_xCoO_2$ phases, $γ-Ca_xCoO_2$ and $γ-Sr_xCoO_2$ (0.15 ≤ x ≤ 0.40) materials have been analyzed. Experimental results demonstrate that the Raman spectra from α, β, and γ -$Na_xCoO_2$ phases have notably different structural features. In general, five Raman active phonons at around 186 $cm^{-1}$ ($E_{1g}$), 470 $cm^{-1}$, 511 $cm^{-1}$, 606 $cm^{-1}$ ($3E_{2g}$), 673 $cm^{-1}$ ($A_{1g}$) appear in $γ-Na_{0.75}CoO_2$; two Raman active phonons at 487 $cm^{-1}$ ($E_g$) and 587 $cm^{-1}$ ($A_{1g}$) appear in $α-Na_xCoO_2$. These results are fundamentally in agreement with the theoretical analysis based on the group factor theory. The Raman spectra of γ-(Sr,Ca)$_x$CoO$_2$ with x = 0.15 and 0.25 show five clear active modes similar with the data from γ-Na$_{0.75}$CoO$_2$, and visible shifts towards higher frequencies appear at $A_{1g}$ and $E_{1g}$ peaks which are essentially connected with the O vibrations. Moreover, the Raman active phonons in γ-(Sr,Ca)$_x$CoO$_2$ materials are found to change apparently with the intercalated cation ordering. These observations suggest that local structural transition in connection with the CoO$_6$ stacking, as well as Sr/Ca intercalation, can result in significant changes in certain phonon modes. Hence, we conclude that the interlayer interaction and local structural features play important roles for understanding the lattice dynamics and physical properties in this layered system. TEM study on samples made by post-chemical treatments, such as Na$_{0.5}$CoO$_2$, could contain different structural phases and complex defects. Therefore, it is worthy to point out that due to the complex structures and various kinds of cation orders in these kinds of materials, special precautions have to be taken for processing and interpreting the experimental data.

## Acknowledgments

We would like to thank Professor N.L. Wang for providing single crystals of Na$_x$CoO$_2$ and Miss G. Zhu for assistance in preparing samples and measuring Raman spectra. We are grateful to Q.M. Zhang and N. Pemberton-Pigott for fruitful discussions and help during manuscript preparation. The work reported here is supported by the 'Outstanding Youth Fund' supported by the National Natural Foundation of China and by the Ministry of Science and Technology of China (973 project No: 2006CB601001).

# Figure Captions

**Fig. 1**: (a) XRD patterns for parent phases of α-NaCoO$_2$, β-Na$_{0.6}$CoO$_2$ and γ-Na$_{0.75}$CoO$_2$. The lattice parameters for all phases are shown for comparison.

(b) Structural models of α-NaCoO$_2$, β-Na$_{0.6}$CoO$_2$ and γ-Na$_{0.75}$CoO$_2$ phases.

**Fig. 2**: Raman spectra of the α-NaCoO$_2$, β-Na$_{0.6}$CoO$_2$ and γ-Na$_{0.75}$CoO$_2$ phases, demonstrating different spectral structures for these specific phases.

**Fig. 3**: (a) Electron diffraction patterns and (b) HRTEM image taken along the [001] zone axis direction of Sr$_{0.35}$CoO$_2$, showing the $3^{1/2}a \times 3^{1/2}a$ superstructure from Sr ordering.

**Fig. 4**: (a) Raman spectra taken from γ-Sr$_x$CoO$_2$ materials with x = 0.15, 0.25, 0.35 and 0.40, showing the evident change of the spectral structures due to the $3^{1/2} \times 3^{1/2}$ superstructure.

(b) Raman spectra taken from γ- Ca$_x$CoO$_2$ materials with x = 0.15, 0.3, 0.35 and 0.40, showing the spectral features arising from cation ordering.

**Fig. 5**: HRTEM and corresponding electron diffraction pattern taken alone [010] zone-axis, indicating the coexistence of different phases observed in the Na$_{0.5}$CoO$_2$ single crystals. The top region is governed mainly by the γ-phase and the bottom area is almost entirely β-phase. C$_1$ is c parameter for γ-phase of about 10.8 Å, C$_2$ is c parameter for β-phase of about 5.7 Å.

Figure 1

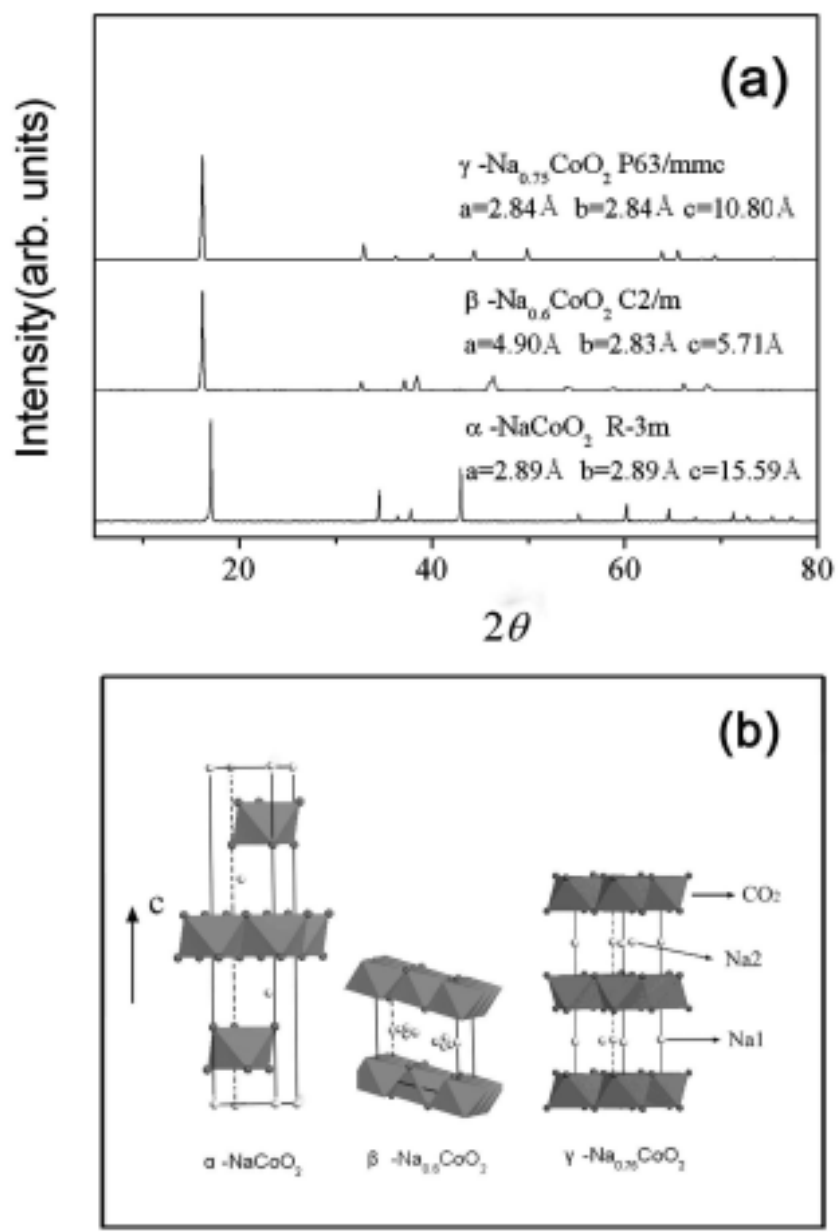

Figure 2

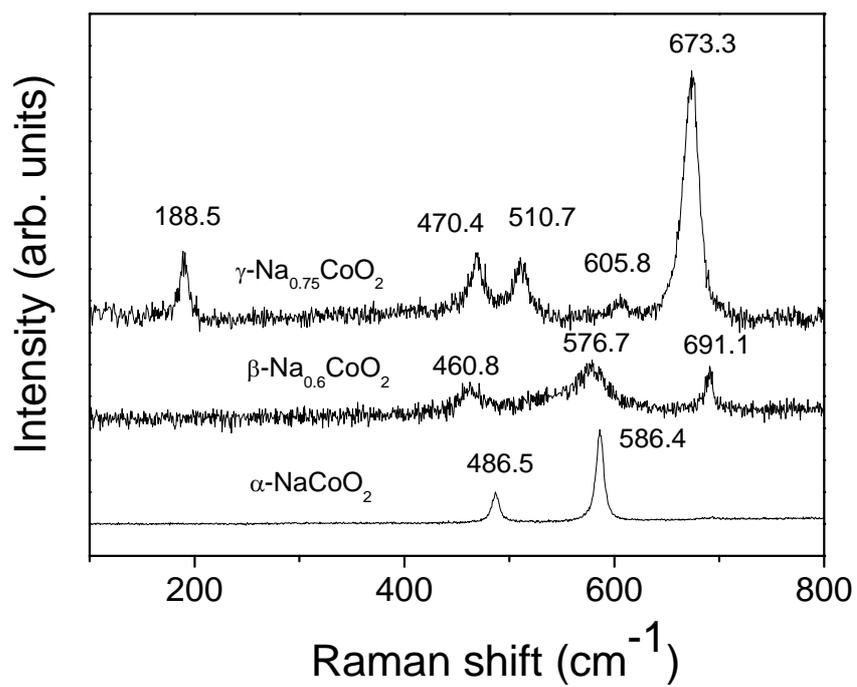

Figure 3

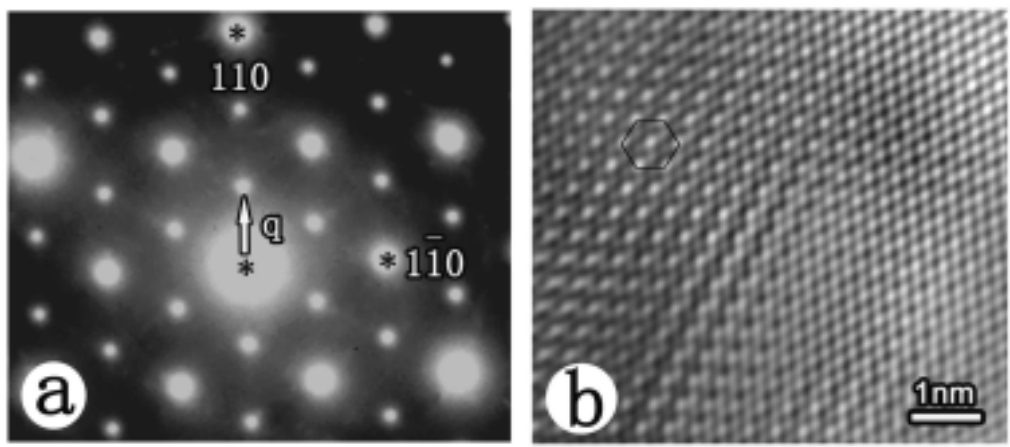

Figure 4

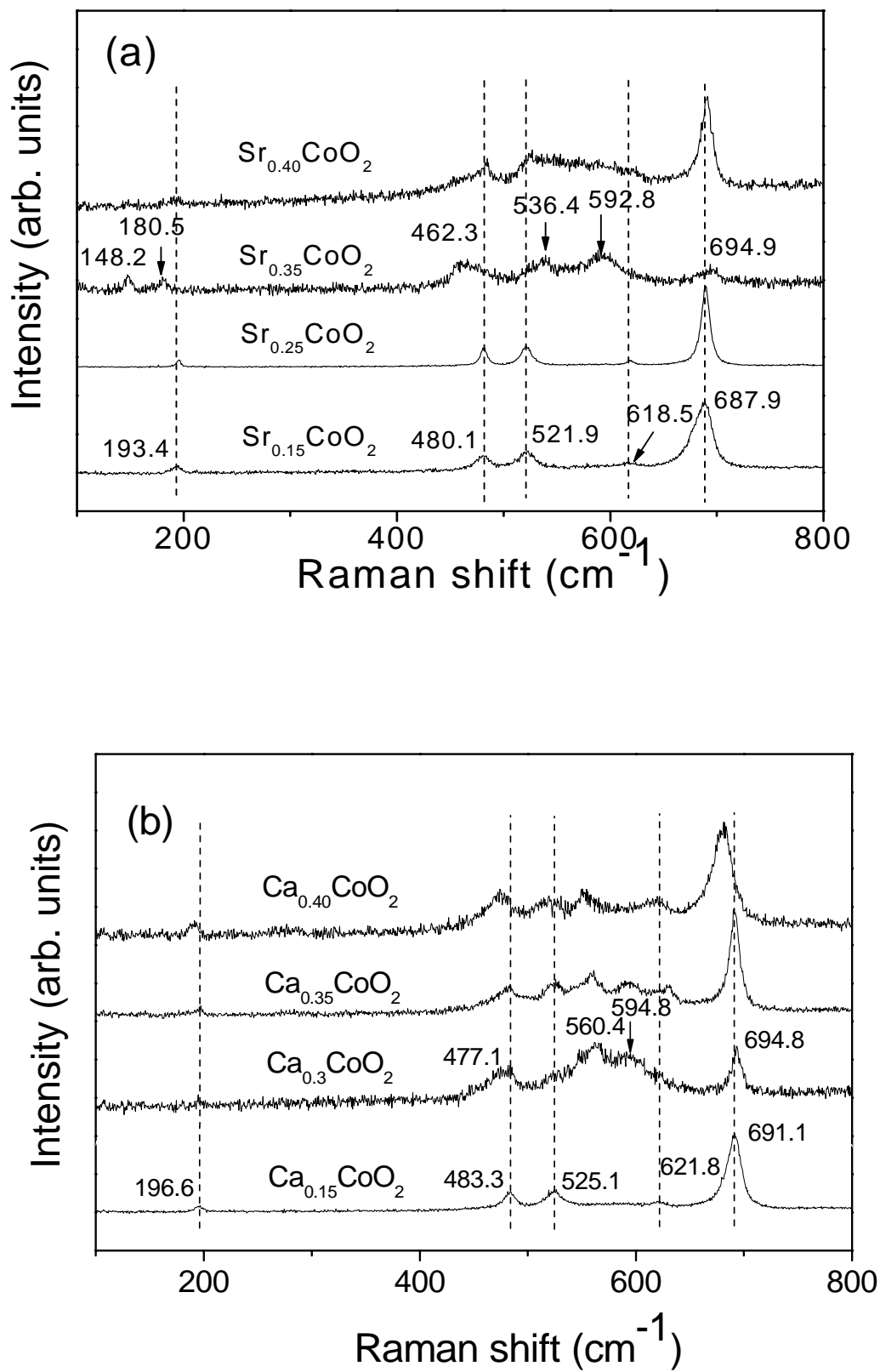

Figure 5

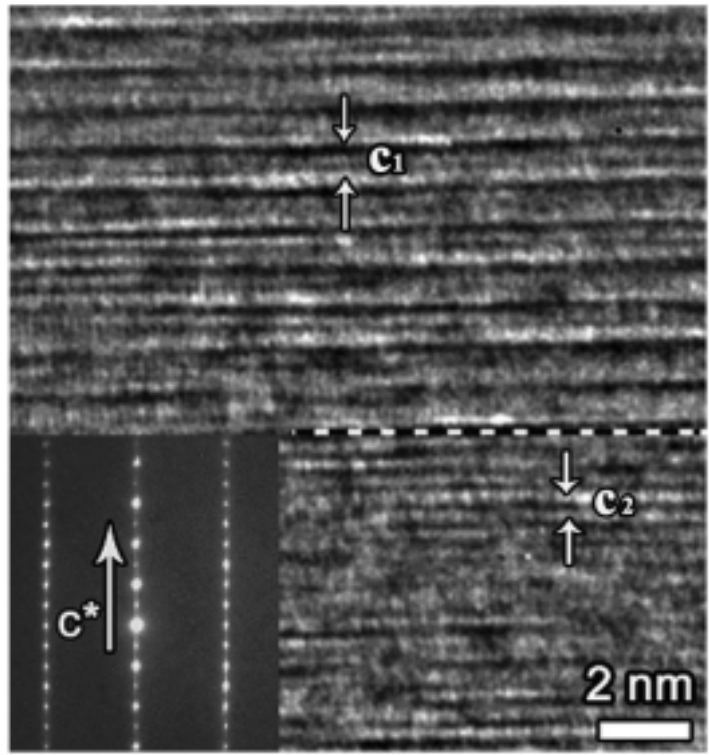